\documentclass[a4paper, twoside, reqno, 12pt, dvips]{amsart}
\RequirePackage{fix-cm} 

\usepackage{fixltx2e}     

\usepackage[english]{babel}
\usepackage[latin1]{inputenc}

\usepackage{eucal}
\usepackage{esint}
\usepackage{amsgen}
\usepackage{amsthm}
\usepackage{xspace}
\usepackage{amssymb}
\usepackage{amsmath}
\usepackage{MnSymbol}
\usepackage{amsfonts}
\usepackage{verbatim}
\usepackage{mathrsfs, dsfont}

\usepackage[nice]{nicefrac}

\usepackage{a4wide}

\usepackage{microtype}
\usepackage[bf, up, hang]{caption}

\usepackage{indentfirst}
\usepackage{graphicx}
\usepackage{subfigure}
\usepackage[section]{placeins}
\usepackage{psfrag}

\usepackage[usenames, dvipsnames, pdf]{pstricks}
\usepackage{epsfig}
\usepackage{pst-grad} 
\usepackage{pst-plot} 

\usepackage{color}
\definecolor{oneblue}{rgb}{0.0, 0.0, 0.85}
\definecolor{darkgrey}{rgb}{0.273, 0.281, 0.30}

\usepackage{xcolor}
\definecolor{lightgray}{gray}{0.9}

\usepackage[colorlinks,
            urlcolor=oneblue,
            linkcolor=oneblue,
            citecolor=oneblue,
            bookmarksopen=false,
            pagebackref]{hyperref}

\usepackage[compact]{titlesec}

\titleformat{\section}{\normalfont\Large\bfseries\sffamily\center\color{darkgrey}}{\thesection.}{0.5em}{}{}
\titleformat{\subsection}{\normalfont\large\bfseries\sffamily\color{darkgrey}}{\thesubsection.}{0.4em}{}{}
\titleformat{\subsubsection}{\normalfont\normalsize\bfseries\sffamily\color{darkgrey}}{\thesubsubsection.}{0.3em}{}{}

\titlespacing*{\section}{1.0em}{1.0em}{0.8em}[0em]
\titlespacing*{\subsection}{1.0em}{1.0em}{0.8em}[0em]
\titlespacing*{\subsubsection}{1.0em}{0.7em}{0.6em}[0em]

\usepackage{titletoc}

\contentsmargin{0.0em}
\dottedcontents{section}[2.5em]{\addvspace{0.45em}\bfseries}{1.0em}{0pc}
\dottedcontents{subsection}[4.5em]{}{2.6em}{0pc}
\dottedcontents{subsubsection}[5.5em]{}{3.0em}{0pc}

\usepackage{fancyhdr}
\usepackage{lastpage}

\newcommand*\Title{Evolution of Random Waves in Finite Depth}
\newcommand*\Authors{D.~Dutykh}

\usepackage{acronym}

\numberwithin{equation}{section}

\newcommand{\up}[1]{$\,^{\mathrm{\small\textsf{#1}}}$} 

\newcommand{\ud}{\mathrm{d}}
\newcommand{\ue}{\mathrm{e}}
\newcommand{\ui}{\mathrm{i}}
\renewcommand{\S}{\varsigma}

\newcommand{\eps}{\varepsilon}

\newcommand{\BFI}{\mathrm{BFI}}

\begin{document}

\title[\Title]{Evolution of Random Wave Fields in the Water of Finite Depth}

\author[D.~Dutykh]{Denys Dutykh$^*$}
\address{University College Dublin, School of Mathematical Sciences, Belfield, Dublin 4, Ireland \and LAMA, UMR 5127 CNRS, Universit\'e de Savoie, Campus Scientifique, 73376 Le Bourget-du-Lac Cedex, France}
\thanks{$^*$ Corresponding author}
\email{Denys.Dutykh@ucd.ie}
\urladdr{http://www.denys-dutykh.com/}

\begin{abstract}
The evolution of random wave fields on the free surface is a complex process which is not completely understood nowadays. For the sake of simplicity in this study we will restrict our attention to the 2D physical problems only (i.e. 1D wave propagation). However, the full Euler equations are solved numerically in order to predict the wave field dynamics. We will consider the most studied deep water case along with several finite depths (from deep to shallow waters) to make a comparison. For each depth we will perform a series of Monte--Carlo runs of random initial conditions in order to deduce some statistical properties of an average sea state.

\bigskip
\noindent \textbf{\keywordsname:} random seas; finite depth; deep water; Euler equations; conformal mapping; Monte--Carlo simulation
\end{abstract}

\maketitle

\tableofcontents

\section{Introduction}

The robust prediction of the wave climate is one of the major scientific and problems nowadays in the oceanography and water wave theory \cite{Hasselmann1976}. On the oceanic scales the statistical description of wave fields is preferred \cite{Hasselmann1962, Hasselmann1980, Komen1996}. The main reasons include, for example, the lack of data to initialize a deterministic model, the need of much higher resolutions (at least a few grid points per wavelengths which becomes practically impossible on large scales), long time simulations to produce a forecast, etc.

The goal of the present study is much more modest. First of all, for the sake of simplicity we will consider the 2D physical problem (1D wave propagation). We will assume that the dynamics is described by the free surface Euler equations \cite{Stoker1958} without any dissipative effects (ideal hydrodynamics) and in the absence of forcing (free decay turbulence). Since the model is deterministic, we will perform numerous Monte--Carlo simulations \cite{Lapeyre2011} of random initial conditions sharing the same statistical characteristics. In this way we hope to be able to draw some statistical conclusions on the behaviour of random wave fields \cite{Shemer2010a}. The main goal of this study is to investigate the effect of water depth on the statistics of the wave field at the free surface. We do not necessarily restrict our attention to narrow banded spectra which appear more often for the mathematical convenience in more theoretical investigations \cite{Dysthe2003}. The computational framework employed here can be used, in principle, for any physically sound initial wave spectra. While most of previous works focussed on the water of infinite depth \cite{Hogan1985, Dysthe2003, Socquet-Juglard2005, Fedele2010, Viotti2013}, A.~\textsc{Toffoli} \emph{et al.} (2009) \cite{Toffoli2009} investigated the statistical properties of random directional wave fields in the finite depth case using the HOS method \cite{Dommermuth1987}. Wave focussing on water of finite depth was studied in \cite{Slunyaev2002}. The transition through the critical regime of vanishing nonlinearity in envelope type equations (when $kh \approx 1.363$) is analyzed in \cite{Slunyaev2005}. For the sake of comparison we consider in this study both cases (finite and infinite depths). The quantities of interest that we consider in this study and follow their evolution are the kurtosis, skewness, BFI index and several others \cite{Janssen2003, Mori2006}. The energy transfer in the finite depth case was studied theoretically earlier by \textsc{Herterich} \& \textsc{Hasselmann} (1980) \cite{Herterich1980}.

The present study is organized as follows. In the following Section~\ref{sec:model} we present the mathematical model along with a brief description of statistical and numerical tools employed in this study. The numerical results are presented in Section~\ref{sec:num}. Finally, the main conclusions and perspectives of this study are outlined in Section~\ref{sec:concl}.

\section{Mathematical model}\label{sec:model}

Consider a two-dimentional (2D) cartesian coordinate system $(x, y)$ with vertical coordinate $y$ pointing upwards. The physical domain is assumed to be bounded below by a horizontal bottom $y = -h$ and above by the free surface $y = \eta(x,t)$. The flow is irrotational and inviscid. The fluid density is set to be unitary without loss of generality. The computational domain $[-\ell, \ell]\times [-h,\; \eta(x,t)]$ is assumed to be periodic in the horizontal direction $x$ in view of application of pseudo-spectral methods. The free surface elevation $\eta(x, t)$ and the velocity potential $\phi(x,y,t)$ are governed by the following system of incompressible and irrotational Euler equations \cite{Stoker1958}:
\begin{equation*}
\begin{array}{lcc}
\left.
\begin{array}{l}
\eta_t    = \displaystyle - {\phi_x} \eta_x + {\phi_y}, \\ 
\\
{\phi_t} = \displaystyle -\frac{1}{2} \left( {\phi_x}^2 + {\phi_y}^2 \right) - g \eta ,
\end{array}
\right\rbrace
\quad &\mbox{at}&
y=\eta(x,t) \\
\\
\nabla^2 \phi=0, \quad &\mbox{for}& \quad   -h < y < \eta(x, t)
\end{array}
\end{equation*}
where $g$ is the gravity acceleration. On the flat bottom the velocity potential satisfies the usual impermeability condition:
\begin{equation*}
  \phi_y = 0, \qquad \mbox{at} \quad y = -h.
\end{equation*}
We note that in the deep water case the depth $h\to +\infty$ and the last condition is replaced by the following one:
\begin{equation*}
 \lim_{y\to-\infty} |\nabla\phi| = 0. 
\end{equation*}

In order to solve this system numerically we employ a pseudo-spectral method similar to one introduced by \textsc{Dyachenko} \emph{et al.} (1996) \cite{Dyachenko1996} (in the infinite depth case), which is based on the conformal transformation that maps the time-dependent flow domain into a half-space \cite{Milewski2010}. Later this approach was generalized for the water of finite depth and even for variable bathymetries \cite{Viotti2013a}. We do not report here the equations resulting from the conformal mapping (these can be found, e.g., in \textsc{Choi} \& \textsc{Camassa} (1999) \cite{Choi1999} or \textsc{Li} \emph{et al.} (2004) \cite{Li2004}. For the time integration we employ the 5\up{th} order Runge--Kutta scheme proposed by \textsc{Dormand} \& \textsc{Prince} (1980) \cite{Dormand1980}.

\subsection{Initial spectrum construction}

All simulations performed in the present study are initialized with a Gaussian sea state. Namely, the Fourier coefficients of the free surface elevation $\eta(x,0)$ and $\phi(x,0)$ are initialized as \cite{Viotti2013}:
\begin{equation*}
\hat{\eta}_{k} = \sqrt{2P_0(k)\ud k} \ue^{\ui k \eps_k}, \qquad 
\hat{\phi}_{k} = -\ui c_k  \hat{\eta}_{k},
\end{equation*}
where $\eps_k$ are independent, uniformly distributed in the interval $[0, 2\pi]$ random phases. We express the potential $\phi_k$ in terms of the free surface displacement $\eta_k$ using the phase velocity $c_k = \omega_k/k$ according to the linear water wave theory ($\omega_k = \sqrt{gk}$ in deep water and $\omega_k = \sqrt{gk\tanh(kh)}$ in finite depth). The function $P_{0}(k)$ is the energy power spectrum chosen in this study to be in the form of a Gaussian:
\begin{equation*}\label{eq:spectrum}
P_0(k) = \frac{\mathcal{P}_{0}}{\sqrt{2\pi\sigma_0^2}} \exp{\left[-\frac{1}{2}\Bigl({k - k_0 \over \sigma_0}\Bigr)^2\right]},
\end{equation*}
where $k_0$ (peak wavenumber) and $\sigma_0$ (spectral width) are some constant positive parameters and $\mathcal{P}_{0}$ is chosen to obtain the prescribed wave amplitude $a_0$.

\subsection{Statistical characteristics}

There are two statistical quantities of particular interest which characterize the wave spectrum \cite{Boccotti2000, Mori2006, Fedele2009, Shemer2010a}:
\begin{itemize}
  \item Kurtosis excess $\kappa := \displaystyle{\frac{\mu_4}{\mu_2^2}} - 3$, which measures the \textit{heaviness} of the spectrum tail.
  \item Skewness $\S := \displaystyle{\frac{\mu_3}{\mu_2^{3/2}}}$, which measures the asymmetry of the spectrum with respect to the mean.
\end{itemize}
These quantities are defined in terms of the free surface elevation moments $\mu_n := \langle \eta^n \rangle$. We note that $\kappa = \S = 0$ for the Gaussian (normal) distribution.

Another important parameter characterizing the spectrum of the surface elevation is the Benjamin--Feir index, $\BFI(t)$. Following Janssen \cite{Janssen2003}, we define this parameter as
\begin{equation*}
  \BFI = \sqrt{2} s\frac{k_{w}}{\sigma_{w}},
\end{equation*}
with spectral width and characteristic wavenumber respectively given by
\begin{equation*}
  \sigma_{w} = \int_0^{+\infty} (k-k_{w})^{2} P\; \ud k\;\; \bigg/ \int_0^{+\infty}  P\; \ud k, \qquad
k_{w} = \int_0^{+\infty}  k P\; \ud k\;\; \bigg/ \int_0^{+\infty}  P\; \ud k,
\end{equation*}
and the averaged wave steepness $s := k_w\; \eta_{\mathrm{rms}}$, where
\begin{equation*}
  \eta_{\mathrm{rms}} \equiv \mu_2^{1/2} = \Bigl\langle 4\int_0^{+\infty}  P\; \ud k \Bigr\rangle^{1/2}.
\end{equation*}
The above definition holds for general spectral shapes, therefore it is suitable for time-evolving spectra. It can be easily checked that initially (at $t = 0$) one has:
\begin{equation*}
  k_w \equiv k_0 \qquad \mbox{and} \qquad \sigma_w \equiv \sigma_0.
\end{equation*}

\section{Numerical results}\label{sec:num}

In this Section we will perform Monte--Carlo simulations of free surface Euler equations in water of infinite ($h = +\infty$) and finite ($h = 2$, $1$, $1/2$) depths\footnote{The depth $h = 1/2$ will correspond approximatively to the shallow water case since the peak wave number $k_0 = 1$ and, thus, $kh = 1/2$.}. The initial condition is generated from the Gaussian power spectrum with random phases. Then, for every initial condition ($M$ Monte--Carlo runs in total) and for every depth ($4$ cases) we simulate the evolution of this wave system using the method of dynamic conformal mappings \cite{Dyachenko1996a} until the final time $T$. The values of all physical and numerical parameters employed in simulations are reported in Table~\ref{tab:params}. These values of parameters produce sea states with the average steepness $s \approx 0.05$ and $\BFI \approx 0.2$. These characteristics can evolve with time and we will discuss it below.

\begin{table}
  \centering
  \begingroup\setlength{\fboxsep}{0pt}
  \colorbox{lightgray}{
  \begin{tabular}{l|c}
  \hline\hline
  Gravity acceleration: $g$ [$\mathsf{m}\,\mathsf{s}^{-2}$] & $1.0$ \\
  Water depth: $h$ [$\mathsf{m}$] & $+\infty$, $2.0$, $1.0$, $0.5$ \\
  Characteristic wavenumber: $k_0$ [$\mathsf{m}^{-1}$] & $1.0$ \\
  Spectral width parameter: $\sigma_0$ [$\mathsf{m}^{-1}$] & $0.29$ \\
  Computational domain half-length: $\ell$ [$\mathsf{m}$] & $100$ \\
  Final simulation time: $T$ [$\mathsf{s}$] & $100.0$ \\
  Initial condition amplitude: $a_0$ [$\mathsf{m}$] & $0.12$ \\
  Number of Fourier modes: $N$ & 8192 \\
  Time step: $\Delta t$ [$\mathsf{s}$] & $0.01$ \\
  Number of Monte--Carlo runs: $M$ & 1000 \\
  \hline\hline
  \end{tabular}}\endgroup
  \caption{\small\em Physical and numerical parameters used for Monte--Carlo simulation of a random wave field evolution under the full free surface Euler dynamics.}
  \label{tab:params}
\end{table}

\begin{figure}
  \centering
  \subfigure[]{\includegraphics[width=0.90\textwidth]{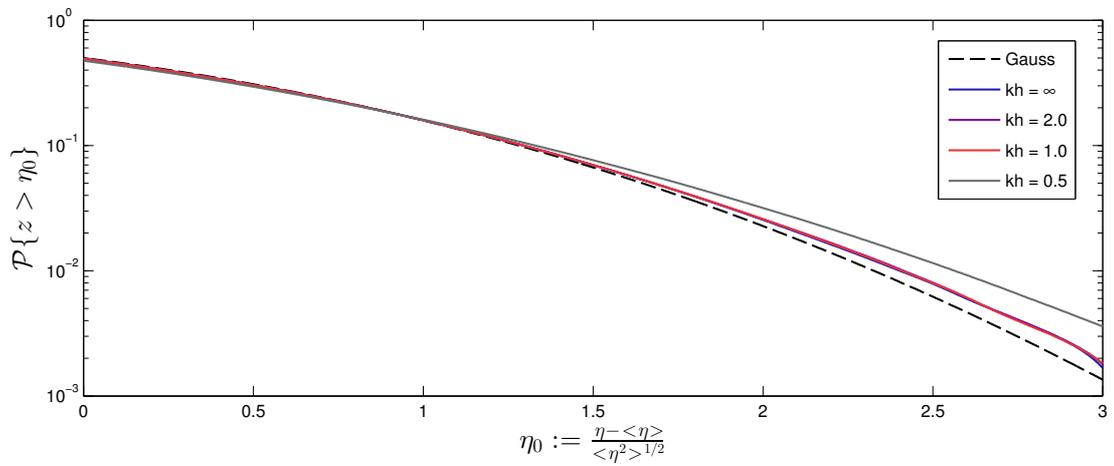}}
  \subfigure[]{\includegraphics[width=0.90\textwidth]{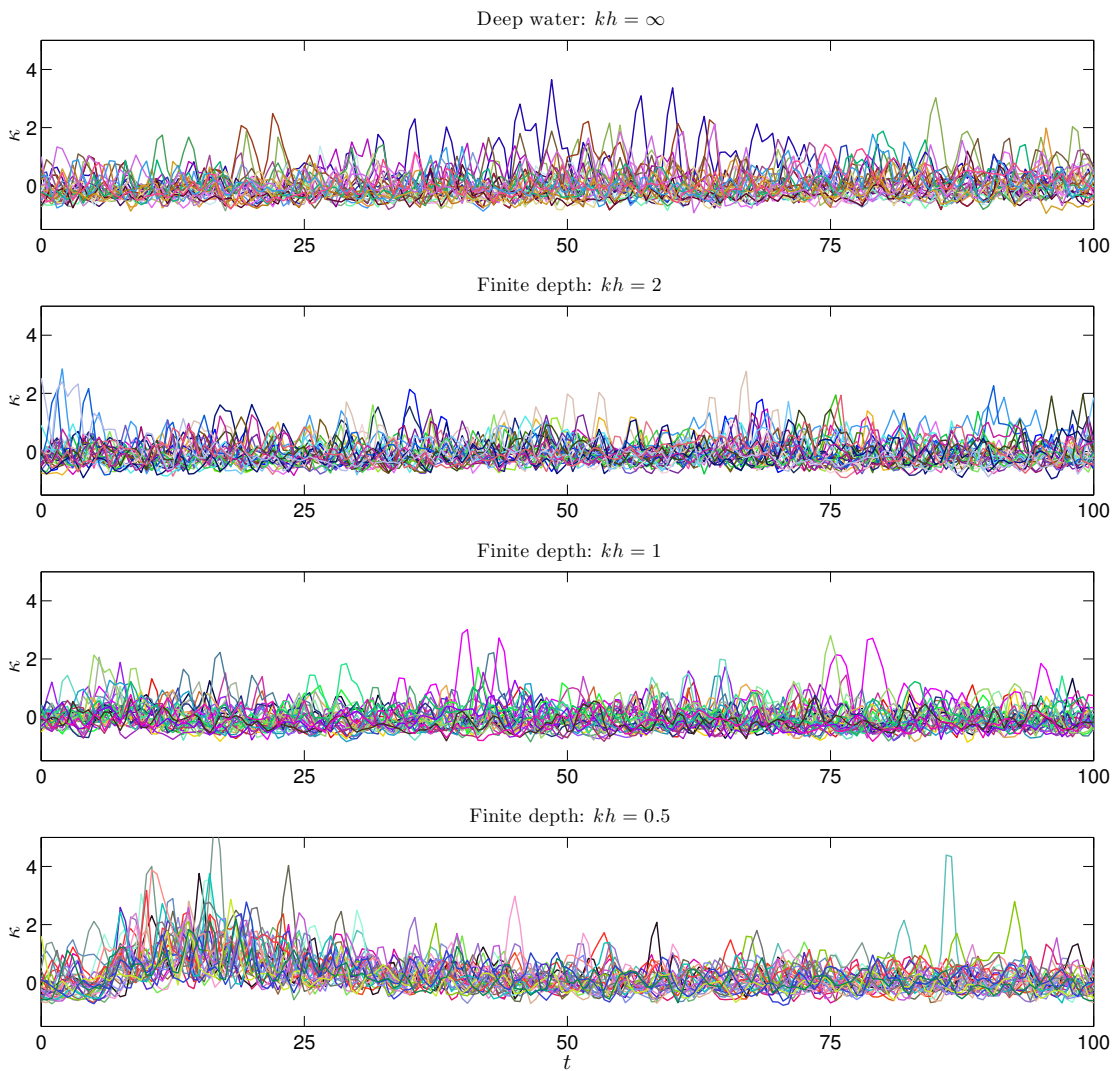}}
  \caption{\small\em (a) Ensemble avergaged excess probability of the free surface elevation at the final simulation time $T$. (b) Sample realizations of the kurtosis excess $\kappa(t)$ for various depths.}
  \label{fig:kurt}
\end{figure}

\begin{figure}
  \centering
  \subfigure[]{\includegraphics[width=0.90\textwidth]{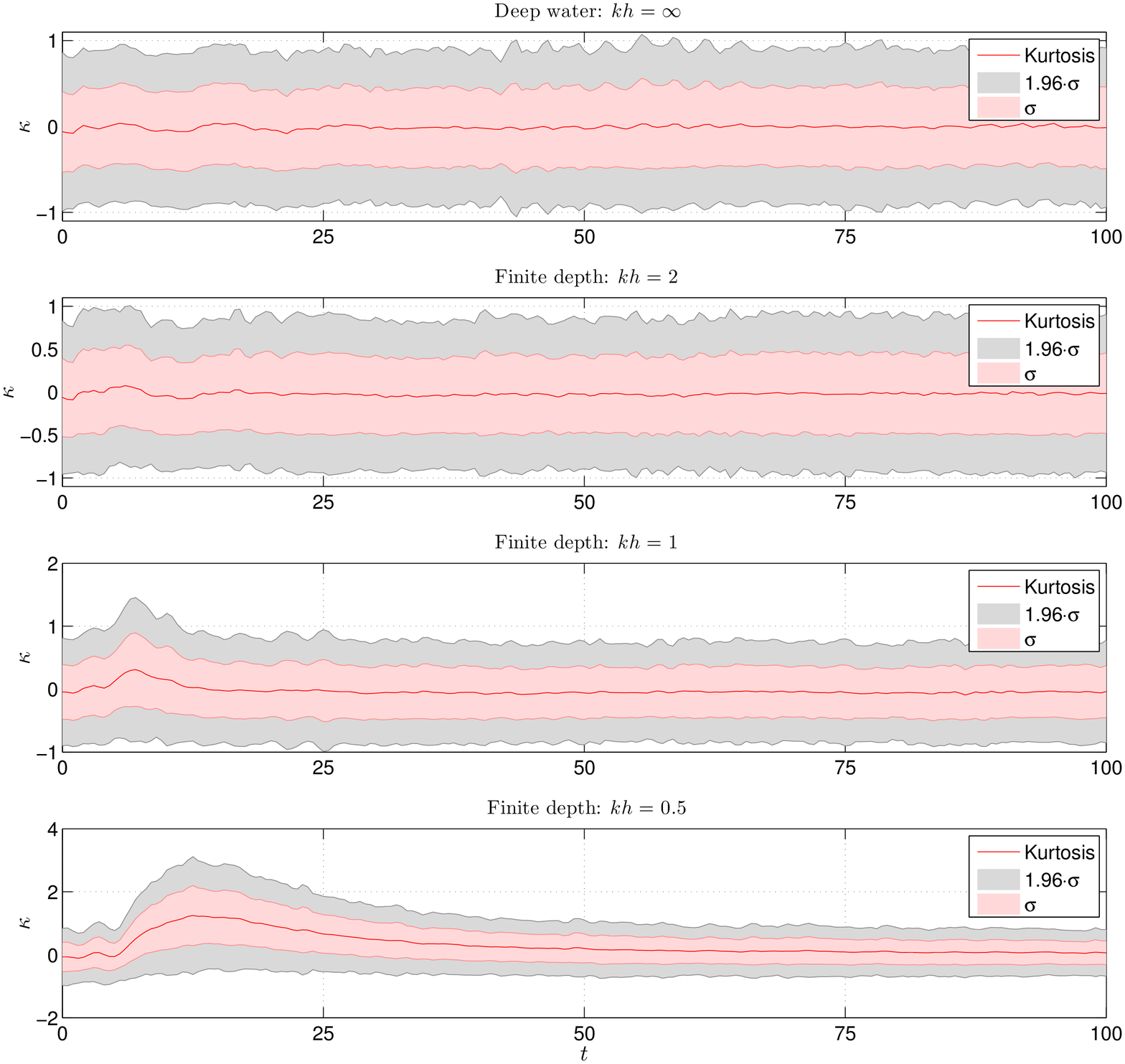}}
  \subfigure[]{\includegraphics[width=0.90\textwidth]{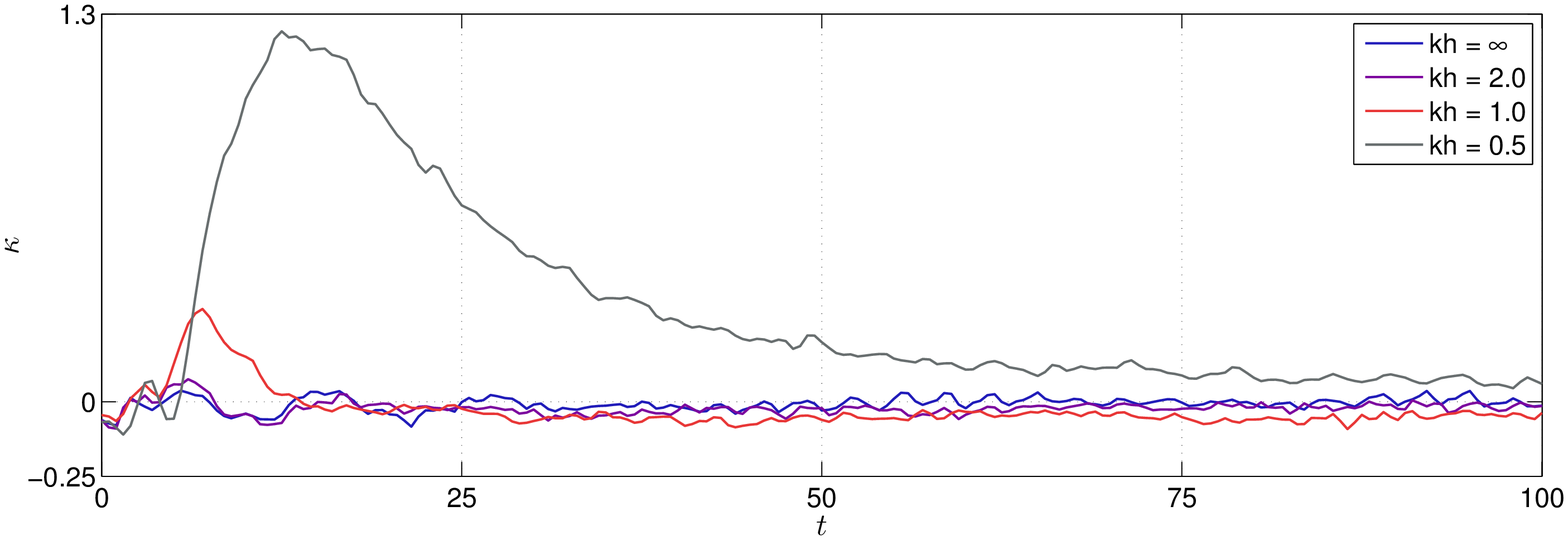}}
  \caption{\small\em (a) Mean value of the kurtosis excess $\kappa(t)$ along with local variance (red area) and 95\% confidence interval (grey area). (b) Ensemble average of the kurtosis excess $\langle\kappa\rangle(t)$ for various depths depicted on the same plot.}
  \label{fig:meankurt}
\end{figure}

\begin{figure}
  \centering
  \subfigure[]{\includegraphics[width=0.90\textwidth]{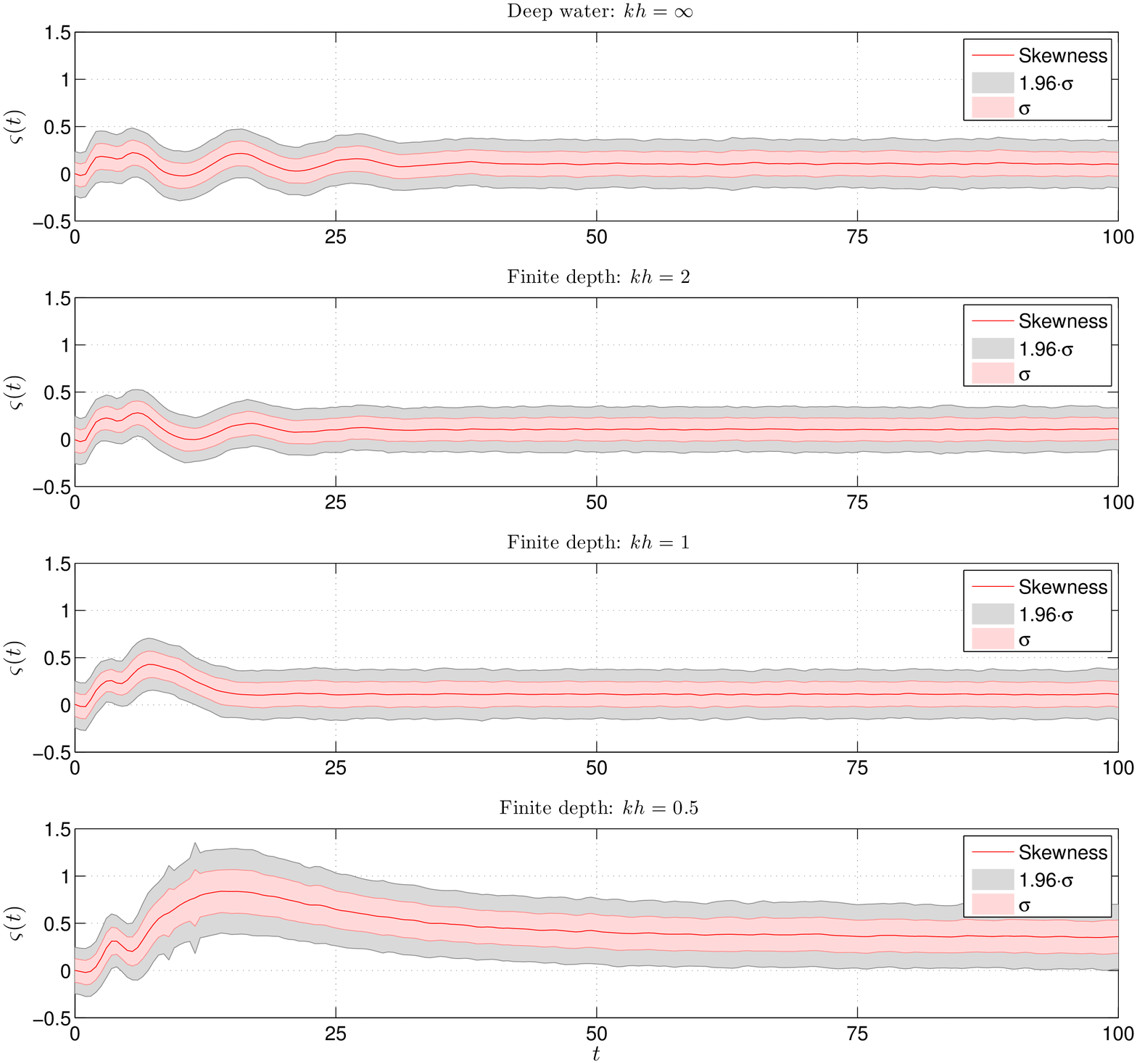}}
  \subfigure[]{\includegraphics[width=0.90\textwidth]{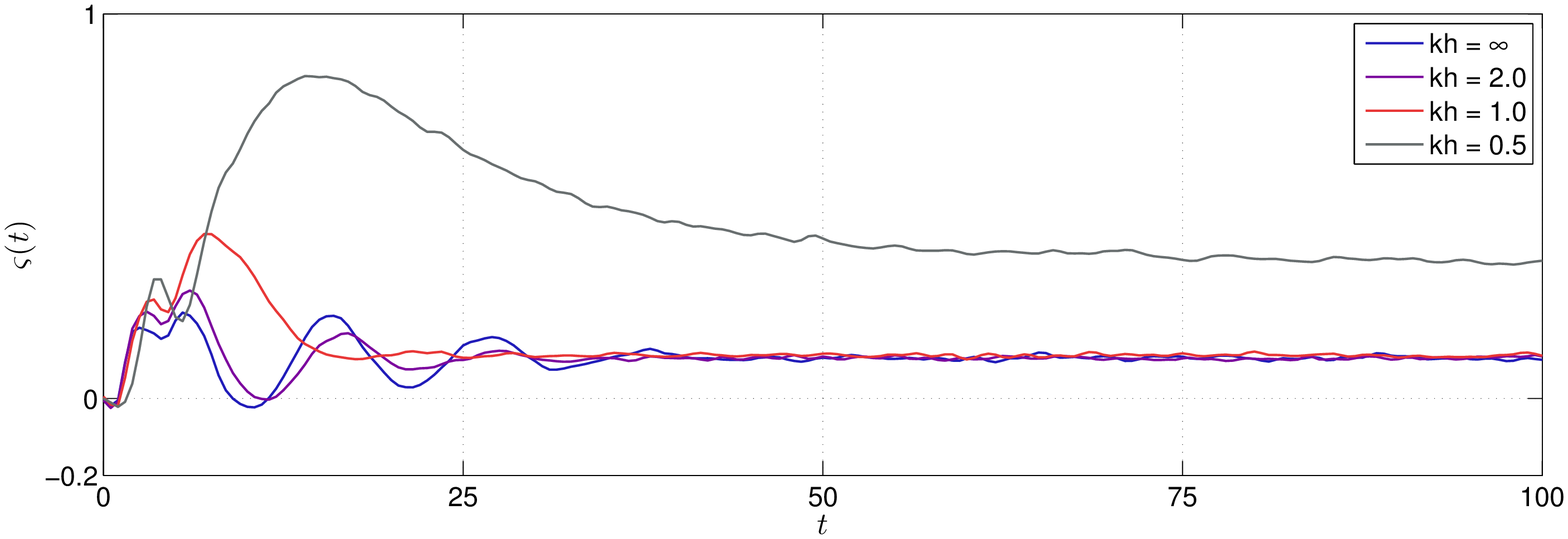}}
  \caption{\small\em (a) Mean value of the skewness $\S(t)$ along with local variance (red area) and 95\% confidence interval (grey area). (b) Ensemble average of the skewness $\langle\S\rangle(t)$ for various depths depicted on the same plot.}
  \label{fig:skew}
\end{figure}

On Figure~\ref{fig:kurt}(a) we plot the excess probability distribution of the (normalized) free surface elevation at the final simulation time $T$. The black dotted line (- - -) represents the classical Gaussian distribution. One can see that there is a clear deviation from the Gaussianity which is very well pronounced in the shallow water case $kh = 1/2$. We constate also that all other \emph{averaged} probability distributions corresponding to $kh = +\infty$, $2$ and $1$ are almost superposed to the graphical resolution. On Figure~\ref{fig:kurt}(b) we show $30$ random realizations of the kurtosis excess $\kappa(t)$ during the simulations for four different depths under consideration. In particular, one can notice important deviations from zero for certain trajectories. The ensemble average of the kurtosis excess is represented on Figure~\ref{fig:meankurt}. In deep waters this statistical quantity averages approximatively to zero. On the other hand, in shallow water the kurtosis excess $\kappa(t)$ shows very important transient deviations and apparently the asymptotic value is above zero. The possibility of large deviations from zero is translated by important variance (red area around the mean). Similar results for the skewness are shown on Figure~\ref{fig:skew}. Again, in shallow waters the skewness remains positive, which indicates that the values above the average are more probable.

\begin{figure}
  \centering
  \subfigure[]{\includegraphics[width=0.90\textwidth]{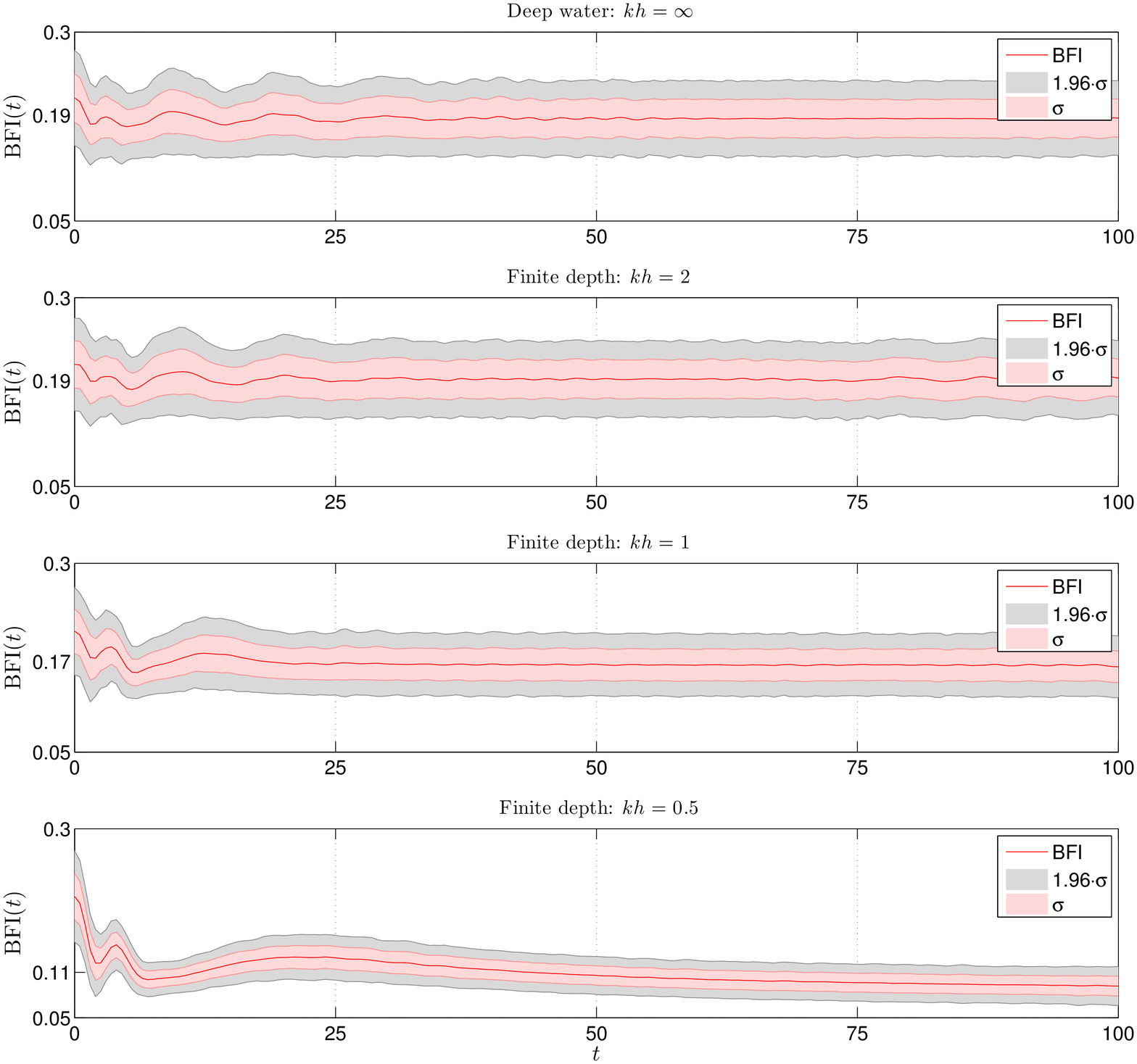}}
  \subfigure[]{\includegraphics[width=0.90\textwidth]{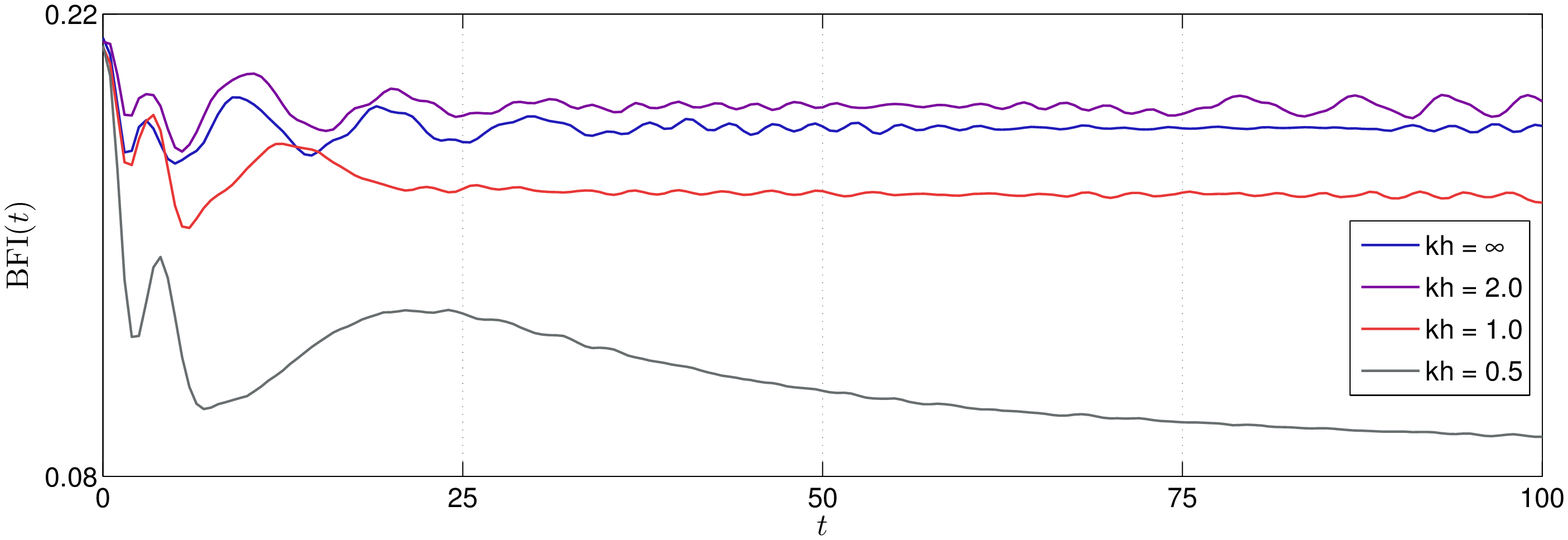}}
  \caption{\small\em (a) Mean value of the Benjamin--Feir index $\BFI(t)$ along with local variance (red area) and 95\% confidence interval (grey area). (b) Ensemble average of the BFI index $\langle\BFI\rangle(t)$ for various depths depicted on the same plot.}
  \label{fig:bfi}
\end{figure}

\begin{figure}
  \centering
  \includegraphics[width=0.90\textwidth]{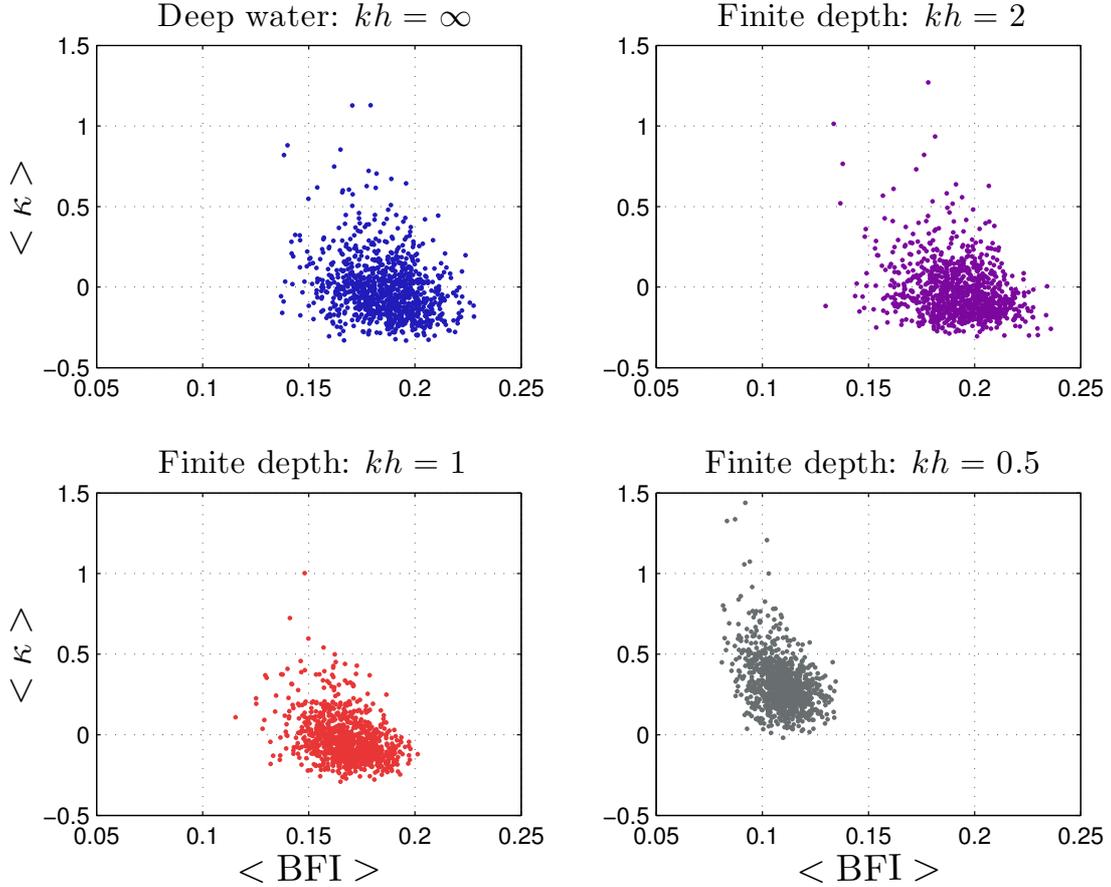}
  \caption{\small\em Time-averaged values of the kurtosis excess $\langle\kappa\rangle$ as the function of the average BFI index $\langle\BFI\rangle$ (in time also) for various depths.}
  \label{fig:KurtBFI}
\end{figure}

The averaged Benjamin--Feir index $\BFI(t)$ is represented on Figure~\ref{fig:bfi}. We remind that all initial conditions share the same statistical characteristics. Nevertheless, in shallow water the stationary averaged BFI index is 50\% lower than in the deep water case. There is an ongoing effort to find a connection between the BFI value and kurtosis \cite{Mori2006}. That is why we produced a scatter plot of these quantities for different values of water depths (see Figure~\ref{fig:KurtBFI}). The scattering is quite important. However, one can see that when the depth is reduced, the whole cloud of points move in certain direction which shows that some dependence between these quantities exist in some very average sense. At the level of a single realization large deviations are possible.

\section{Conclusions and perspectives}\label{sec:concl}

In this study we presented preliminary results on the effect of the water depth on the statistical characteristics of the quasi-deterministic wave field. Namely, we showed that statistical properties may vary significantly when the depth is decreased from deep to rather shallow waters. Even in deep waters our numerical results show an interesting result that the ensemble average of the kurtosis is equal to (statistical) zero even if the variance is rather important, showing that large deviations are to be expected with high probability. On the other hand, in shallow waters the variance of kurtosis is significantly lower, but the mean value can make important transient excursions. The skewness remains positive in average, which indicates that the values above the average are more probable as expected. We note also that the asymptotic value of skewness is higher in the shallow water case ($k_0d = 1/2$).

Concerning the perspectives for the nearest future, we would like first of all to perform more simulations with various values of the BFI index (here we reported only results for $\BFI \approx 0.2$). On the other hand, larger domains (and thus bigger number of waves) and longer simulation times (to allow for the development of wave train instabilities) are also needed to decrease the statistical error of reported results.

\section*{Acknowledgements}

This publication has emanated from research conducted with the financial support of European Research Council under the research project ERC-2011-AdG 290562-MULTIWAVE. The author wishes to acknowledge the SFI/HEA Irish Centre for High-End Computing (ICHEC) for the provision of computational facilities and support under the project ``\textit{Numerical simulation of the extreme wave run-up on vertical barriers in coastal areas}''. D.~\textsc{Dutykh} is grateful to F.~\textsc{Fedele} for fruitful discussions on random waves.

\addcontentsline{toc}{section}{References}
\bibliographystyle{abbrv}
\bibliography{biblio}

\end{document}